\documentclass[letterpaper,english,aps,prb,reprint,superscriptaddress]{revtex4-2}
\usepackage[T1]{fontenc}
\usepackage{color}
\usepackage{mathrsfs}
\usepackage{amsmath}
\usepackage{amssymb}
\usepackage{makeidx}
\makeindex
\usepackage{graphicx}

\makeatletter

\pdfpageheight\paperheight
\pdfpagewidth\paperwidth

\usepackage{hyperref}
\hypersetup{
    colorlinks = true,
	allcolors = blue	
}
\usepackage{times}

\makeatother

\usepackage{babel}
\begin{document}
\preprint{\textcolor{red}{\printindex\bibliographystyle{plainnat}
\bibliography{zotero_bib,ref}
}APS/123-QED}
\title{Altermagnetism-Induced Parity Anomaly in Weak Topological Insulators}
\author{Yu-Hao Wan}
\affiliation{International Center for Quantum Materials, School of Physics, Peking University, Beijing 100871, China}

\author{Qing-Feng Sun}
\thanks{Corresponding author: sunqf@pku.edu.cn.}
\affiliation{International Center for Quantum Materials, School of Physics, Peking University, Beijing 100871, China}
\affiliation{Hefei National Laboratory, Hefei 230088, China}

\begin{abstract}
We demonstrate that introducing altermagnetism on the surface of a
weak topological insulator (TI) results in the emergence of a single
massless Dirac fermion, exhibiting a parity anomaly.
To explore the transport properties induced by this parity anomaly,
we propose an effective
two-dimensional (2D) lattice model to describe the weak TI surface.
This model captures both the energy spectrum and spin texture of the
weak TI surface while reducing computational complexity. We show that
the weak TI surface hosts a half-integer chiral edge current under
the influence of altermagnetism. Additionally, in the presence of
decoherence, the Hall conductance attains a half-quantized value.
Layer-resolved calculations from a 3D slab model further confirm that
surface altermagnetism drives the surface Hall conductance to transition
to $e^{2}/2h$, aligning with calculation from the 2D effective lattice model.
Our findings establish a link between altermagnetism and quantum anomalies,
positioning weak TIs as a potential platform for investigating the
parity anomaly without a net magnetic moment.
\end{abstract}
\maketitle

\section{Introduction}

Parity anomaly is a concept from high-energy physics, where a two
dimensional massless Dirac fermion with parity symmetry couples to
a gauge field, inevitably generating a small mass term that breaks
parity symmetry through a regularization process\citep{redlich1984gaugenoninvariance,redlich1984parityviolation,semenoff1984condensedmatter}.
The realization of parity anomaly in condensed matter systems is a
fundamental problem and has been the subject of extensive research\citep{bottcher2019survival,chang2013experimental,zhang2017anomalous,mciver2020lightinduced,lu2018topology,ozawa2019topological,jotzu2014experimental,zhou2022transport,xu2014observation,tokura2019magnetic,ning2023robustness,yang2014classification,fu2022quantum}.
To achieve parity anomaly in condensed matter, the system must host
a single Dirac cone. As a consequence of the parity anomaly, a single
Dirac cone results in several significant observable effects. For
example, in electronic systems, the parity anomaly leads to a half
quantized Hall effect, as predicted by the anomaly-induced Chern-Simons
theory\citep{haldane1988modelfor,semenoff1984condensedmatter,lapa2019parityanomaly,burkov2019diracfermion,zhou2022transport}.
In Majorana systems, where a massless Majorana fermion couples to
a gravitational field, the parity anomaly gives rise to a gravitational
Chern-Simons term, resulting in a quarter quantized thermal Hall conductance\citep{furusaki2013electromagnetic,sekine2021axionelectrodynamics,sato2016majorana,wang2011topological,wan2024quarterquantized}.

The realization of parity anomaly in condensed matter systems was
first proposed by Haldane \citep{haldane1988modelfor}, who suggested
that by introducing a staggered magnetic flux and adjusting the on-site
energy difference between the A and B sublattices in a honeycomb lattice,
the system could achieve parity anomaly when one of the two valleys
is fine-tuned to close while the other remains gapped. However, Haldane's
proposal is difficult to realize in real materials due to the challenges
in precisely tuning the staggered magnetic flux. Nevertheless, it
has been successfully implemented in artificial system such as photonic
\citep{liu2020observation,lu2017observation} and acoustic system\citep{ding2019experimental}.
Another approach to realizing the parity anomaly in condensed matter
systems is through the use of the surface states of three dimensional
topological insulators (3D TIs)\citep{qi2008topological,fu2007topological,qi2011topological2}.
Recently, experiments involving semi-magnetic 3D TIs have achieved this,
successfully realizing the parity anomaly \citep{mogi2022experimental}.
In these experiments, a magnetic moment was introduced on one surface
of a 3D TI, which opened a gap in the Dirac cone on that surface,
while leaving the Dirac cone on the opposite surface gapless. This
experimental setup led to the observation of the half quantized Hall
effect, a direct consequence of the parity anomaly in the system.

While the parity anomaly has been experimentally realized on the surface
of strong 3D TIs, its realization in weak 3D TIs remains a significant
challenge. This difficulty arises from the fundamental difference
in their topological properties: strong TIs, with nontrivial $\mathbb{Z}_{2}$
topology, host a single Dirac cone on their surfaces, whereas weak
TIs exhibit an even number of Dirac cones due to their trivial bulk
$\mathbb{Z}_{2}$ topological invariant\citep{fu2007topological}.
Introducing ferromagnetic ordering in a weak TI would gap out all
surface Dirac cones simultaneously, preventing the system from hosting
the single Dirac cone necessary for the realization of the parity
anomaly. Therefore, achieving the parity anomaly in weak 3D TIs requires
a different mechanism from that used in strong TIs.

Altermagnetic magnetism, a distinct type of magnetic ordering, is
believed to exist in various materials and can induce momentum-dependent
spin splitting without net macroscopic magnetization\citep{krempasky2024altermagnetic,mazin2021prediction,vsmejkal2022beyondconventional}.
This unique property has made altermagnetism a topic of considerable
research interest, both theoretically and experimentally, with studies
exploring its connections to superconductivity \citep{vsmejkal2022emerging,ouassou2023dcjosephson,chakraborty2024zerofield,cheng2024fieldfree,banerjee2024altermagnetic,giil2024superconductoraltermagnet,sun2023andreev},
topological phenomena \citep{li2024creation,fernandes2024topological},
magnetic multipoles \citep{bhowal2024ferroically,fernandes2024topological},
and anomalous Hall effect\citep{vsmejkal2022anomalous}.
%However, the connection between altermagnetism and quantum anomalies remains
%unexplored.

In this work, we propose that altermagnetism can be employed to achieve
the parity anomaly in weak 3D TIs. By introducing altermagnetism in proximity
to the surface of weak TIs, one of the two gapless Dirac cones can
be gapped out, leaving a single Dirac cone on the surface, which allows
for the realization of the parity anomaly.
To investigate the surface transport properties
induced by the parity anomaly, we constructed
a 2D lattice model for the surface of the weak 3D TI by introducing an
anisotropic Wilson mass term. This 2D model effectively captures the
linear dispersion of the Dirac cones at the $\Gamma$ and $Y$ points
on the weak 3D TI surface. Additionally, the spin texture distribution
obtained from the 2D lattice model closely matches the spin texture
observed on the surface of the weak TI. Calculations of the equilibrium
current reveal that the presence of a single Dirac cone leads to a
half-integer chiral edge current. Furthermore, using the nonequilibrium
Green's function method, we computed the Hall conductance under dephasing.
The results show that, under dephasing, the half-quantized Hall conductance
appears. Despite the absence of a net magnetic moment, the system
still exhibits half quantized Hall conductance due to the parity anomaly.
This finding establishes a profound connection between altermagnetism
and quantum anomaly and offers a novel platform for studying the parity
anomaly.

The rest of the paper is organized as follows. In Sec. II, we introduce
the model Hamiltonian of the weak 3D TI and explain how a single Dirac
cone is realized by introducing altermagnetism on the surface. In
Sec. III, we present the 2D lattice model for the surface states of
the weak 3D TI and validate its accuracy by comparing the band structure
and spin texture with those of the weak 3D TI. In Sec.
IV, using the 2D lattice model developed in Sec. III and considering
the proximity of altermagnetism, we calculate the equilibrium current
and the Hall conductance under dephasing. In Sec. V, we compute the
layer-resolved Hall conductance distribution of the weak
TI in the presence of surface-proximal altermagnetism.
Finally, a brief summary is provided in Sec. VI.

\begin{figure}
\begin{centering}
\includegraphics[scale=0.4]{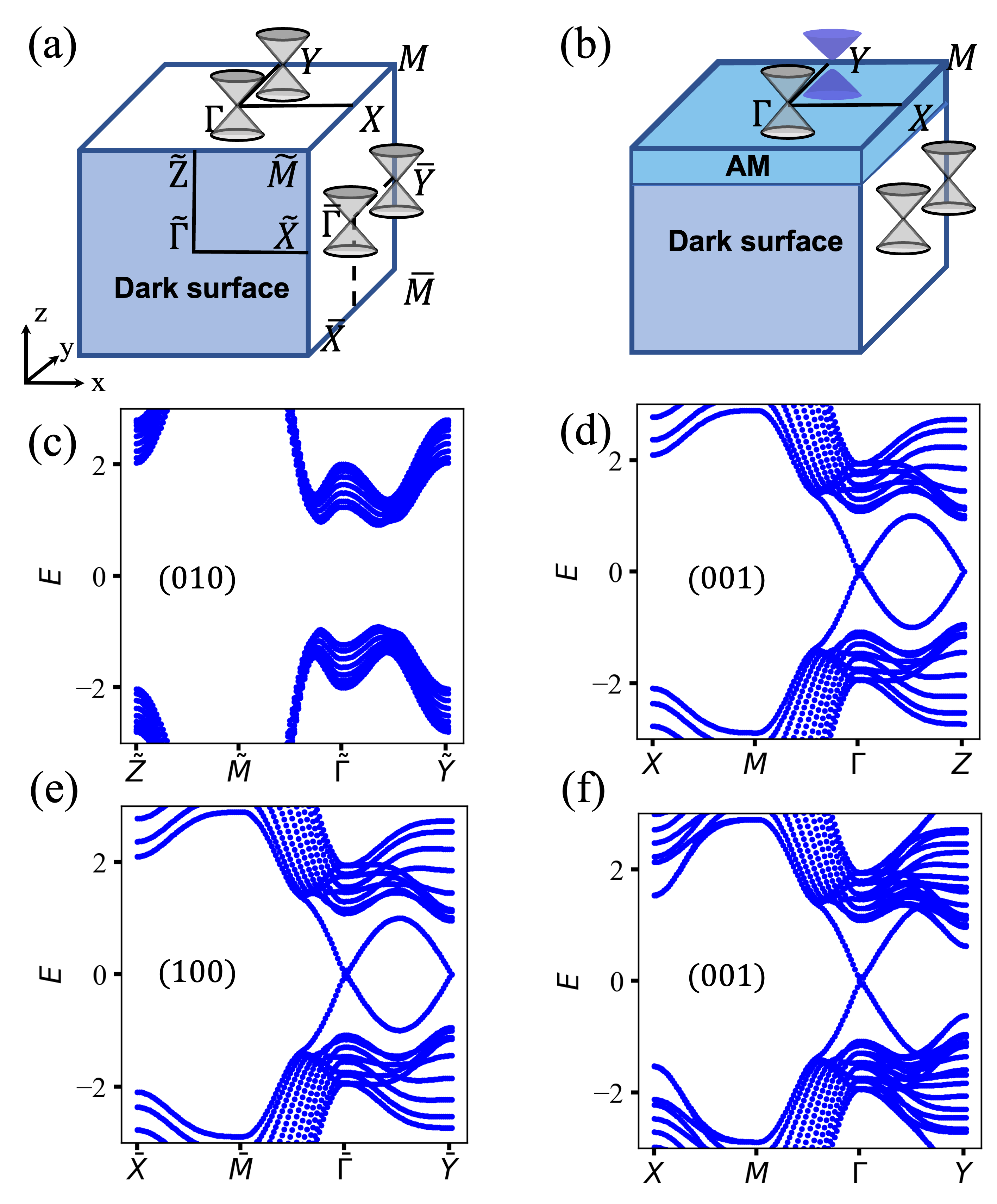}
\par\end{centering}
\caption{\label{fig:1}(a) Schematic of the weak TI, with the dark 
surface (010) marked in blue. 
Two gapless Dirac cones are present on the (100) and (001) surfaces. 
(b) On the (001) surface with altermagnetism, the
Dirac cone at the $Y$ point is gapped. (c-e) Band structures of the
slab models stacked along the (010), (001), and (100) directions,
respectively, without altermagnetism. (f) Energy spectrum of the slab
model stacked along the (001) direction with altermagnetism applied
on the surface, with an altermagnetic strength of $J=0.5$.}
\end{figure}

\section{Mode Hamiltonian of weak TI}

We begin with the tight-binding Hamiltonian for the 3D TI in a cubic
lattice\citep{wan2024magnetizationinduced}:
\begin{eqnarray}
\mathcal{H}&=&\left(m-\sum_{i}2B_{i}\right)s_{0}\sigma_{z} \nonumber\\
&& +\sum_{i}\left(2B_{i}s_{0}\sigma_{z}\cos k_{i}+As_{i}\sigma_{x}\sin k_{i}\right),\label{eq:weak3d}
\end{eqnarray}
where $m$, $B_i$, and $A$ are the model parameters, and $k_{i}$
is the momentum with $i=x,y,z$. Here, $s_{i}$ and $\sigma_{i}$
are the Pauli matrices in the spin and orbital spaces, respectively.
To position the system in the weak topological phase, we set $A=1$,
$B_{y}=0.2$, $B_{x}=B_{z}=1$, and $m=2$, unless otherwise specified.
Under these parameters, the system has a $\mathbb{Z}_{4}$ index of
$(0,010)$\citep{fu2007topological,fu2007topological2}. This $\mathbb{Z}_{4}$ index signifies that the phase is adiabatically connected to QSH states stacked along the (010) direction, indicating
the presence of an even number of Dirac cones on the (100) and (001)
surfaces, while the (010) surface remains a gapped dark surface (See
Fig.\ref{fig:1}(a)). This behavior is similar to that of the realistic
weak TI $\textrm{Zr\ensuremath{_{}}Te\ensuremath{_{5}}}$\citep{zhang2021observation},
which features even Dirac cones on the side surfaces.

To investigate the surface energy spectrum, we construct 10-layers
slab models stacked along $l$-directions. The Hamiltonian is represented
as follows:
\begin{equation}
H_{\text{slab}}^{l}=\begin{pmatrix}H_{\text{lay}_{1}} & D & 0 & 0\\
D^{\dagger} & H_{\text{lay}_{2}} & D & 0\\
0 & D^{\dagger} & \dots & D\\
0 & 0 & D^{\dagger} & H_{\text{lay}_{10}}
\end{pmatrix}\label{eq:salb}
\end{equation}
where $l$ represents the stacking direction, which can be one of
$x,y,z$. The Hamiltonian for each layer is given by:
\begin{align*}
H_{\text{lay}}(k_{m},k_{n}) & =A \sigma_{x} s_{m} \sin k_{m}+ A \sigma_{x} s_{n} \sin k_{n} \\
 & +\left[m+2B_{m}(\cos k_{m}-1)+2B_{n}(\cos k_{n}-1)\right]\sigma_{z} s_{0},
\label{Hlay}
\end{align*}
where $k_{m}$ and $k_{n}$ represent the lattice momenta in the $m$
and $n$ directions, respectively, which are perpendicular to $l$.
The interlayer hopping is given by: $D=\frac{A}{2i}\sigma_{x} s_{l}+B_{l}\sigma_{z} s_{0}$.

In Figures \ref{fig:1}(c-e), we present the band spectrum derived
from the slab models. The band spectrum for the slab model stacked
along the (100) direction is gapped, indicating a ``dark
surface" in that direction. In contrast, the band spectra
for the slab models stacked along the (001) and (010) directions exhibit
two gapless Dirac cones located at the $\Gamma$/$\bar{\Gamma}$ and
$Y/\bar{Y}$ points. 

Subsequently, we introduce altermagnet coupled 
on the (001) surface to open the gap for the Dirac cones located at the $Y$, 
as depicted in Fig. \ref{fig:1}(b).
The proximity of the altermagnet modifies the surface Hamiltonian through the following term:
\begin{equation}
H_{\text{AM}}(k_{x},k_{y})=J(\cos k_{x}-\cos k_{y})\sigma_{0} s_{z}(\delta_{l,1}+\delta_{l,10}), \label{HAM}
\end{equation}
with $J$ the altermagnetic strength.
To facilitate the examination of the influence of surface altermagnetism
on the energy band structure, symmetric magnetization is applied on
both the top and bottom surfaces. Figure \ref{fig:1}(f) illustrates
the band structure of the slab model stacked along the (001) direction
after the introduction of the altermagnetic term. It is evident that
the Dirac cone at the $Y$ point becomes gapped, whereas the Dirac
cone at the $\Gamma$ point remains gapless. This suggests that altermagnet
can give rise to a surface with a single gapless Dirac cone in a weak
TI.

\section{the 2D Lattice Model for the Surface States of
the Weak 3D TI}

To study the transport properties of the surface states, directly
computing based on the Hamiltonian derived from a 3D lattice model
can be computationally expensive. Achieving high-precision results
often requires working with an extremely large Hilbert space. Although
weak 3D TIs exhibit an insulating bulk state, their surface states
near the Dirac nodes can be effectively described by a massless 2D
Dirac Hamiltonian at low energies. Compared to 3D lattice models,
2D lattice models significantly reduce computational complexity while
preserving key physical properties. In the case of strong 3D TIs,
well-established 2D lattice models already exist\citep{zhou2017twodimensional}.
However, for weak 3D TIs, there is still a lack of an efficient lattice
model to describe their surface states.

Next, we construct a 2D lattice model for the surface of a weak TI,
specifically considering the (001) surface. We start from a Hamiltonian
of massless 2D Dirac fermion : $H_{Dirac}=\widetilde{A}(k_{x}\sigma_{x}+k_{y}\sigma_{y})$,
where$\widetilde{A}$ is a parameter corresponding to the Dirac fermion
velocity. We discretize it on a square lattice by replacing $k_{i}\rightarrow\frac{1}{a}\sin(k_{i}a)$,
where $\ensuremath{a}$ is the lattice constant, and we set $a=1$
for simplicity. Thus, the lattice Hamiltonian of the 2D Dirac fermion
becomes:
\begin{equation}
H_{Dirac}^{Lattice}=\widetilde{A}[\sin(k_{x})s_{x}+\sin(k_{y})s_{y}]. \label{HDirac}
\end{equation}

Due to the fermion doubling problem in lattice models of massless
Dirac fermions\citep{redlich1984gaugenoninvariance}, there are Dirac
cones at all four high-symmetry points $\Gamma$, $M$, $X$, and
$Y$. However, for the surface of a weak 3D TI described by Eq.(\ref{eq:weak3d}),
we expect only two Dirac cones located at the $\Gamma$ and $Y$ points.
To address this, we introduce an anisotropic Wilson mass term:
\begin{equation}
H_{W}^{aniso}(k_{x},k_{y})=2\widetilde{B_{x}}(1-\cos(k_{x}))s_{z}. \label{HW}
\end{equation}

Thus, the 2D lattice model for the surface of the weak 3D TI can be
written as:
\begin{equation}
H_{surf}=H_{Dirac}^{Lattice}+H_{W}^{aniso}.\label{H2dsurf} 
\end{equation}
The anisotropic
Wilson term vanishes as $k_{x}$ approaches zero in a quadratic manner,
which opens gaps at the Dirac cones at the $M$ and $X$ points while
leaving the Dirac cones at the $\Gamma$ and $Y$ points gapless.
This can be easily understood from the energy spectrum. The energy
spectrum of $H_{surf}$ can be expressed as:
\[
E_{k_{x},k_{y}}^{2}=\widetilde{A}^{2}\sum_{i=x,y}\sin^{2}(k_{i})
+(4\widetilde{B_{x}})^{2}\sin^{4}(k_{x}/2).
\]
As $\ensuremath{(k_{x},k_{y})\rightarrow\Gamma\text{ or }Y}$, $\ensuremath{E^{2}}$
approaches $\widetilde{A}^{2}\sum_{i=x,y}q_{i}^{2}+\widetilde{B_{x}}^{2}q_{x}^{4}$,
where $q_{x}$ and $q_{y}$ represent small deviations from the momenta
near the $\Gamma$ and $Y$ points. This implies that the spectrum
remains linearly dispersive at the $\Gamma$ and $Y$ points. However,
as $\ensuremath{(k_{x},k_{y})\rightarrow X\text{ or }M}$, $\ensuremath{E^{2}}$
approaches $\widetilde{A}^{2}\sum_{i=x,y}q_{i}^{2}+16\widetilde{B_{x}}^{2}$,
with the Wilson term opening a gap of size $8\widetilde{B_{x}}$ at
the $M$ and $X$ points.

\begin{figure}
\begin{centering}
\includegraphics[scale=0.45]{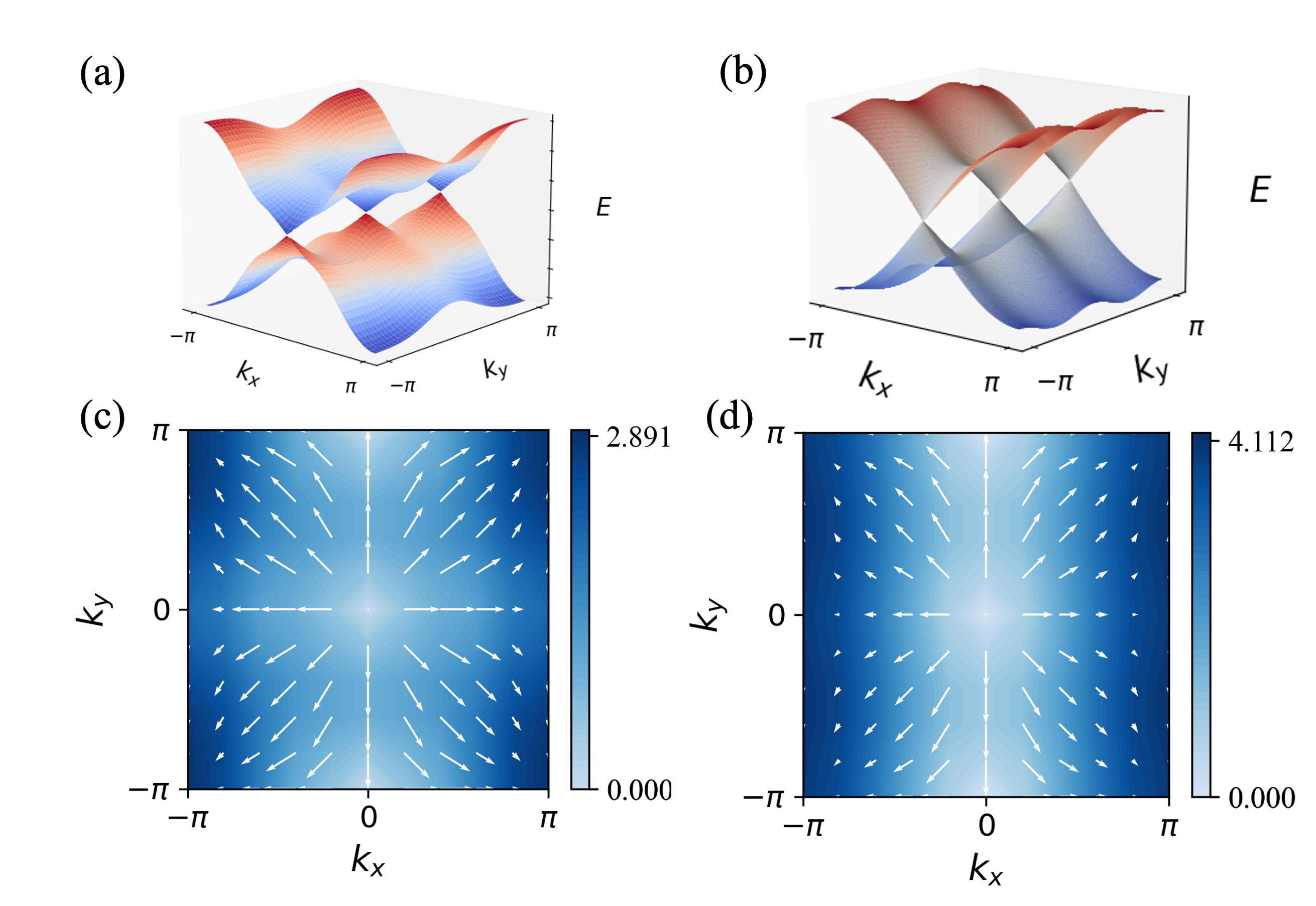}
\par\end{centering}
\caption{\label{fig:2}(a) Energy spectrum of the (001) surface of the weak
3D TI. (b) Energy spectrum of the 2D lattice model with $\widetilde{A}=1$
and $\widetilde{B_{x}}=1$. (c, d) Spin textures of the positive-energy
bands from (a) and (b), respectively, where the color represents the
magnitude of the energy.}
\end{figure}

To demonstrate the 2D lattice model in Eq.(\ref{H2dsurf}) effectively captures the properties
of the weak TI surface, we compare the surface spectrum of a weak
3D TI and the spectrum given by the 2D lattice model. Figures
\ref{fig:2}(a) and \ref{fig:2}(b) show the band spectra of the (001)
surface of the weak 3D TI and the band structure of 2D lattice model,
respectively. In both cases, we observe two gapless Dirac cones at the
$\Gamma$ and $Y$ points.

Furthermore, we compare the spin textures of the two models. Figure
\ref{fig:2}(c) shows the spin expectation value $\left\langle \hat{\mathbf{S}}\right\rangle $
for the surface states of the weak 3D TI in the Brillouin zone, where
$\left\langle \hat{\mathbf{S}}\right\rangle =\sum_{i=x,y}\langle\psi_{k}|s_{i}\sigma_{x}|\psi_{k}\rangle$,
with $|\psi_{k}\rangle$ representing the Bloch states of the surface
states with positive energy at momentum $k$. From the spin texture,
it is evident that the Dirac cones at the $\Gamma$ and $Y$ points
exhibit opposite chirality, corresponding to Berry phases of $+\pi$
and $-\pi$, respectively. Similarly, Figure \ref{fig:2}(d) displays
the spin texture obtained from the 2D lattice model.
%, where the spin
%operator is represented by Pauli matrices $s_{i}$. 
The spin texture
derived from the lattice model closely resembles that of the weak
3D TI, particularly near the $\Gamma$ and $Y$ points, reflecting
the same chirality as observed in the 3D model's calculations.

In constructing the effective surface model for weak TIs, the introduction of an anisotropic Wilson mass term effectively captures the underlying physics. A weak TI with a \(\mathbb{Z}_4\) index of \((0,010)\) can be viewed as a stack of quantum spin Hall (QSH) states along the (010) direction\citep{fu2007topological,fu2007topological2}, where the surface states can be considered as being composed of quasi-one-dimensional edge states derived from the QSH layers. Consequently, the origin of the surface states in weak TIs can be attributed to the topological properties of the stacked QSH layers. Therefore, when constructing the lattice Hamiltonian for the surface, we introduce a Wilson mass term along the in-plane direction of the QSH layers to capture the influence of the nontrivial bulk topology of the QSH states on the boundaries.

\begin{figure}
\begin{centering}
\includegraphics[scale=0.4]{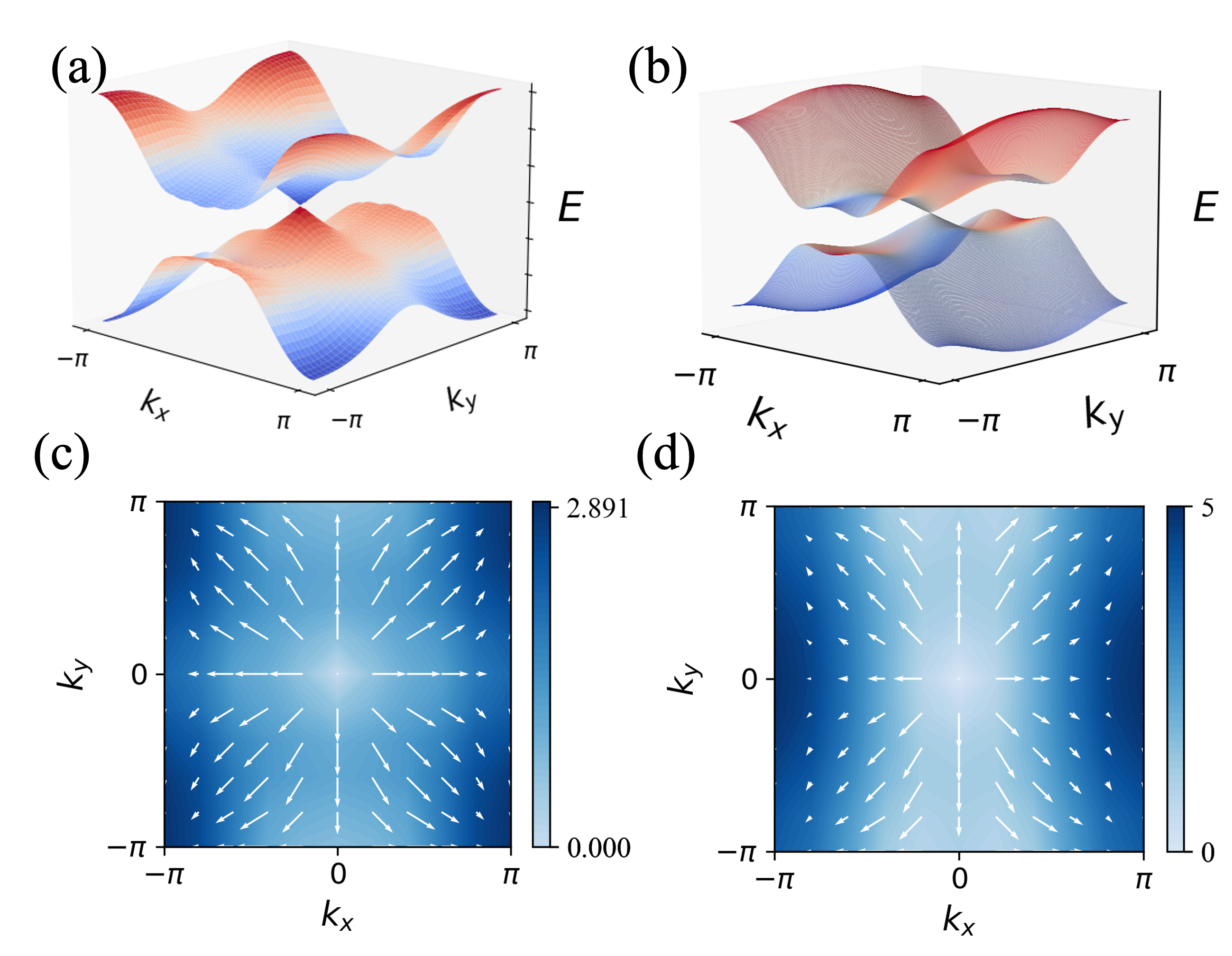}
\par\end{centering}
\caption{\label{fig:3}(a) Energy spectrum of the (001) surface of the weak
3D TI under altermagnetism, with $J=0.5$. (b) Energy spectrum of
the 2D lattice model under altermagnetism, with $\widetilde{A}=1$,
$\widetilde{B_{x}}=1$, and $J=0.5$. (c, d) Spin textures of the
positive-energy bands from (a) and (b), respectively, where the color
represents the magnitude of the energy.}
\end{figure}

Next, we introduce altermagnetic term, $J(\cos k_{x}-\cos k_{y})s_{z}$, to 2D lattice model.
Under the influence of altermagnetism, the energy spectrum becomes
$E_{k_{x},k_{y}}^{2}=\widetilde{A}^{2}\sum_{i=x,y}\sin^{2}(k_{i})
+[(4\widetilde{B}_{x})\sin^{2}(k_{x}/2)+J(\cos k_{x}-\cos k_{y})]^{2}.$
In this case, as $\ensuremath{(k_{x},k_{y})\rightarrow Y}$, $\ensuremath{E^{2}}$
approaches $\widetilde{A}^{2}\sum_{i=x,y}q_{i}^{2}+4J{}^{2}$. The
Dirac cone at the $Y$ point is gapped due to the introduction of
altermagnetism, resulting in a gap of $4J$, while leaving only a
single gapless Dirac cone at the $\Gamma$ point. Figures \ref{fig:3}(a)
and (b) display the energy spectra for the weak TI surface and the
2D lattice model under the influence of altermagnetism, both showing
a gapless Dirac cone at the $\Gamma$ point. Furthermore, the introduction
of altermagnetism does not affect the distribution of the spin texture.
Figures \ref{fig:3}(c) and (d) illustrate the spin textures for the
weak TI surface and the 2D lattice model, respectively. In both cases,
a clockwise winding spin texture is observed around the $\Gamma$
point, corresponding to a Berry phase of $\text{\ensuremath{\pi}}$.

\section{Equilibrium Flow and Half-Quantized Hall Conductance}

\begin{figure}
\begin{centering}
\includegraphics[scale=0.37]{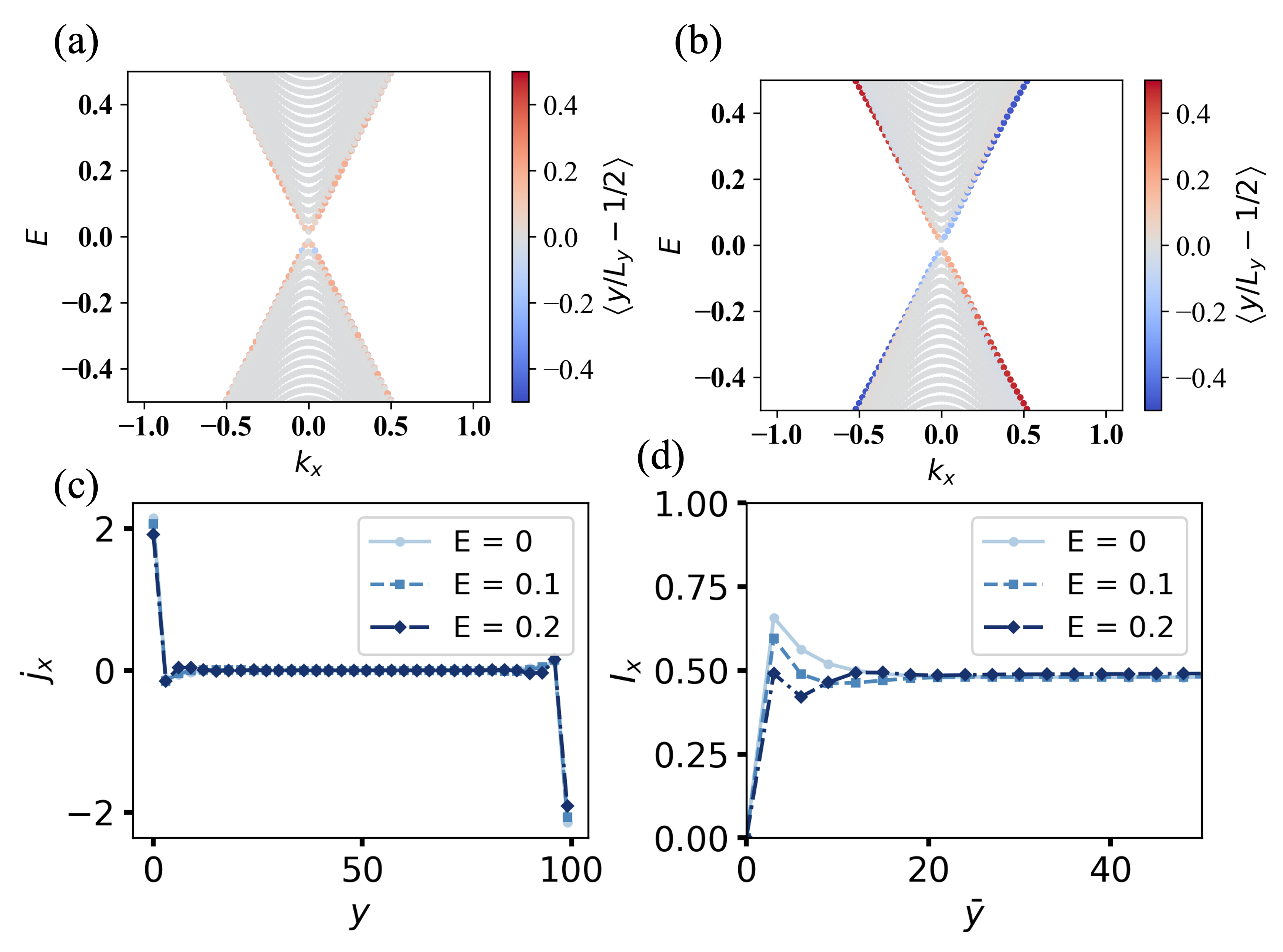}
\par\end{centering}
\caption{\label{fig:4}(a, b) Band structure of the nanoribbon, with colors
indicating the average displacement relative to the center for each
Bloch state, $\langle x/L_{x}-1/2\rangle$. The calculation is performed
with $L_{y}=100$, and for $J=0$ in (a) and $J=0.5$ in (b). 
(c) Distribution of the equilibrium current along the $y$-direction at
different energies. (d) Variation of the current flux as the summation
range $\overline{y}$ increases.}
\end{figure}

Building on the 2D lattice model for the surface of the weak 3D TI, we
now proceed to investigate the transport properties of the surface
under the influence of altermagnetism. 
This approach allows us to focus specifically on 
the surface states' behavior while also significantly
reducing the computational complexity compared to a full 3D model.
To begin, we derive the Hamiltonian for the nanoribbon by discretizing
$H_{surf}$ in Eq.(\ref{H2dsurf}) along the y-direction while maintaining translational
symmetry in the x-direction:
\begin{eqnarray}
H_{quasi-1D} & = &\sum_{k_{x},y}\left(c_{k_{x},y}^{\dagger}T_{y}\,c_{k_{x},y+1}\,+\,\text{h.c.}\right) \nonumber\\
 & +& c_{k_{x},y}^{\dagger}\left[\widetilde{A}\sin(k_{x})\,s_{x}+2\widetilde{B}_{x}\left(1-\cos(k_{x})\right)s_{z} \right. \nonumber\\
 && \left. +J\cos k_{x}\,s_{z}\right]c_{k_{x},y}
\end{eqnarray}
where $c_{k_{x},y}^{\dagger}$ and $c_{k_{x},y}$ are creation and
annihilation operators at position $\ensuremath{y}$ with momentum
$k_{x}$ and $T_{y}=-i\widetilde{A}\,s_{y}-\frac{J}{2}s_{z}$. 

Figures \ref{fig:4}(a) and \ref{fig:4}(b) show the energy spectrum of the
single gapless Dirac cone at the $\Gamma$ point 
for a nanoribbon along the x-direction with $J=0$ and $J=0.5$, respectively. 
The color in the plots represents the average displacement relative to
the center for each Bloch state, $\left\langle y/L_{y}-1/2\right\rangle $
with $L_{y}$ the width of the nanoribbon.
Without altermagnetism $(\ensuremath{J=0})$, the energy spectrum
does not reveal any obvious chiral features [see Fig. \ref{fig:4}(a)].
However, after introducing altermagnetism $(\ensuremath{J=0.5})$,
the states on the left and right sides of the energy spectrum
with opposite velocities 
locate at the upper and lower sides of the nanoribbon 
(exhibiting the red and blue colors in Fig. \ref{fig:4}(b)), 
respectively, indicating the emergence of chirality. This behavior reflects the breaking of time-reversal symmetry at the surface due to altermagnetic effects, leading to the separation of counterpropagating modes. Despite lacking a net magnetic moment, altermagnetism can still induce and control chiral edge states.

To further illustrate the altermagnetism-induced chiral nature on
the surface of the weak 3D TI, we further calculated the equilibrium
current distribution for the case of $J=0.5$, with the equilibrium
current $j_{x}$ denotes as\citep{gong2023halfquantized}:
\[
\ensuremath{\begin{aligned} & j_{x}(E,y)\\
 & =-\frac{e}{\pi h}\int_{-\pi}^{\pi}\operatorname{Im}\operatorname{Tr}\left[\frac{\partial H\left(k_{x}\right)}{\partial k_{x}}G_{k_{x}}^{R}(E,y,y)\right]dk_{y},
\end{aligned}
}
\]
where $G_{k_{y}}^{R}\left({E,y,y^{'}}\right)$ is the retarded Green's
function for $H(k_{x})$, written as $\ G_{k_{x}}^{R}(E,y,y')=\langle y|\left(E-H(k_{x})+i\eta\right)^{-1}|y'\rangle$,
with $\eta=0.1$. 
Fig. \ref{fig:4}(c) shows the distribution of the equilibrium current $j_{x}$
in the y-direction for different energies. The existence of counterpropagating
currents at the upper and lower boundaries confirms the chiral nature
of the system. Furthermore, as shown in Fig. \ref{fig:4}(d), the
current flux at different energies, $\ensuremath{{I_{x}}\left(E,{\bar{y}}\right)=\int_{0}^{\bar{y}}{dy}{j_{x}}\left(E,{y}\right)}(\bar{y}<L_{y}/2)$
gradually converges to 1/2 as $\overline{y}$ increases, demonstrating
the presence of the half quanzited equilibrium current, a hallmark of systems exhibiting parity anomaly\citep{zou2022halfquantized}.

\begin{figure}
\begin{centering}
\includegraphics[scale=0.43]{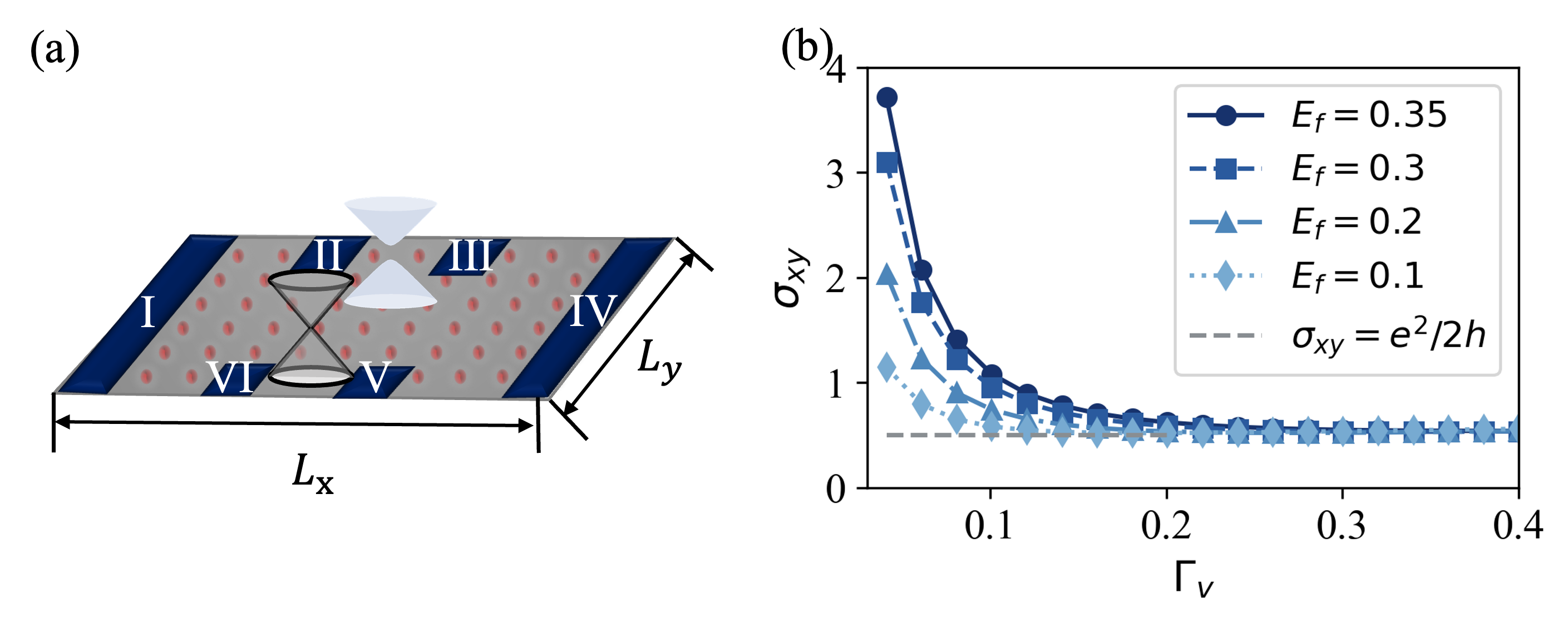}
\par\end{centering}
\caption{\label{fig:5}(a) Illustration of the 2D lattice model of the surface
of a weak 3D TI, with the attached ports shown in dark blue. Due to the
effect of altermagnetism, only a single gapless Dirac cone exists
on the surface. The red balls represent the virtual leads. (b) Hall
conductance as a function of dephasing strength $\Gamma_{v}$ for
different Fermi energy levels $E_{F}$.}
\end{figure}

In fact, this half-integer equilibrium current is related to the half
quantized Hall conductance of the system. Experimentally, the Hall
conductance can be measured using a six-terminal transport Hall-bar setup.
For numerical simulations, we proceed by discretizing Hamiltonian
$H$ in both the $x$- and $y$-directions, with the lengths $L_{x}$
and $L_{y}$ both set to 180 in the following calculations. We then
construct a six-terminal Hall-bar device (see Fig. \ref{fig:5}(a)),
where terminals I and IV are current leads, and terminals II, III,
V, and VI serve as voltage probes. The dephasing effect is simulated
by introducing Buttiker virtual leads between each pair of lattice
sites\citep{buttiker1986fourterminal,xing2008influence,buttiker1988symmetry},
represented by the red spheres in Fig. \ref{fig:5}(a).

Using the Landauer-B\"{u}ttiker formula, the current at terminal $n$
at zero temperature is given by\citep{long2011quantum,lambert1993multiprobe,zhou2024dissipative}:
\begin{equation}
I_{n}=\sum_{m\neq n}\frac{e^{2}}{h}(T_{mn}V_{n}-T_{nm}V_{m}),\label{eq:landuer}
\end{equation}
where $V_{m}$ is the voltage at lead $m$, and $T_{nm}(E_{F})$
denotes the transmission coefficient at energy $E_F$ from lead $m$
to lead $n$. The transmission coefficient is expressed as: $T_{nm}(E)=\operatorname{Tr}[\Gamma_{n}\mathbf{G}^{R}\Gamma_{m}\mathbf{G}^{A}]$,
where $\Gamma_{m/n}=i(\Sigma_{m/n}^{r}-\Sigma_{m/n}^{a})$ is the
line-width function for leads $m/n$, with $\Sigma_{m/n}^{r}$ and
$\Sigma_{m/n}^{a}$ being the retarded and advanced self-energies 
due to the coupling of the leads $m/n$, respectively. 
The retarded Green's function $\mathbf{G}^{R}(E)$ is
given by: $\text{\ensuremath{\mathbf{G}^{R}}(E)=\ensuremath{\left[\mathbf{G}^{A}\right]^{\dagger}}=\ensuremath{\left[(E+i\eta)\mathbf{I}-\mathbf{H}-\sum_{n}\Sigma_{n}^{r}\right]^{-1}}}$.
For the virtual leads, the self-energy is $\Sigma_{n}^{r}=-\frac{i}{2}\Gamma_{v}I_{n}$,
where $\Gamma_{v}$ is the dephasing strength \citep{xing2008influence},
and $I_{n}$ is a $2\times2$ identity matrix. Using Eq. (\ref{eq:landuer}) 
and assuming the voltage of the left (I) and right (IV) 
terminals being $V/2$ and $-V/2$,
we can calculate the currents and voltages at all terminals. Due to
current conservation, we have $|I_{1}|=|I_{4}|$, and the Hall resistance
$R_{xy}$ and longitudinal resistance $R_{xx}$ are obtained as:
\begin{equation}
R_{xy}=\frac{V_{2}-V_{6}}{I_{2}},\quad R_{xx}=\frac{V_{2}-V_{3}}{I_{2}}.
\end{equation}
The resistivities are given by $\rho_{xy}=R_{xy}$ and $\rho_{xx}=\frac{R_{xx}}{l_{1}/L_{y}}$,
where $l_{1}$ is the distance between leads II and III, set to 10
in the calculations. Finally, the Hall conductance is determined by:
\begin{equation}
\sigma_{xy}=\frac{\rho_{xy}}{\rho_{xy}^{2}+\rho_{xx}^{2}}.
\end{equation}

Figure \ref{fig:5}(b) shows the variation of the calculated Hall
conductance as the dephasing strength $\Gamma_{v}$ increases at different
Fermi energy levels. As the dephasing strength increases, the Hall
conductance rapidly decays to $\ensuremath{e^{2}/2h}$. This half-quantized
Hall conductance is independent of the Fermi energy, a characteristic
consistent with the parity anomaly observed in semi-magnetic TIs\citep{zhou2022transport,mogi2022experimental}.
It is important to consider the dephasing effect because, in real
experiments, sample sizes often extend to several hundred micrometers,
far exceeding the decoherence length, thus necessitating the inclusion
of dephasing in the analysis.

\section{layer-resolved Hall conductance}

After discussing the transport properties derived from the 2D lattice
model for the surface of the weak 3D TI, we now turn our attention to
the layer-resolved Hall conductance in a weak 3D TI\citep{varnava2018surfaces,bianco2011mapping,rauch2018geometric}.
This layer-resolved Hall conductance can be obtained by calculating
the local Chern number, where the Hall conductance in layer $l$ is
related to the local Chern number between layers through the relation
$\sigma_{xy}(l)=C_{z}(l) \frac{e^{2}}{h}$, with $C_{z}(l)$ representing
the local Chern marker along the $z$-direction.

We add an altermagnetic term $\ensuremath{J(\cos k_{x}-\cos k_{y})\sigma_{0}\otimes s_{z}\delta_{l,1}}$
to the first layer of the 10-layer slab model in Eq. (\ref{eq:salb}).
Afterward, we compute the local Chern marker projected onto layer
$l$, denoted as $C_{z}(l)$. The expression for the local Chern marker
is given by \citep{varnava2018surfaces}:
\begin{equation}
C_{z}(l)=\frac{-4\pi}{\mathscr{A}}\operatorname{Im}\frac{1}{N_{k}}\sum_{k}\sum_{v,v',c}X_{vck}Y_{v'ck}^{\dagger}\rho_{vv'k}(l),
\end{equation}
where $\mathscr{A}$ represents the unit cell area, $N_{k}$ is the
number of $k$-points, and $\rho_{vv'k}(l)$ is the projection matrix
onto the orbitals within layer $l$, summing over all orbitals belonging
to that layer. $X$ and $Y$ denote the position operators along the
$x$- and $y$-directions, respectively. The matrix elements of the
position operator, denoted as:
\begin{equation}
X\left(Y\right)_{vc\mathbf{k}}=\left\langle \psi_{v\mathbf{k}}\right|x\left(y\right)\left|\psi_{c\mathbf{k}}\right\rangle =\frac{\left\langle \psi_{v\mathbf{k}}\right|i\hbar v_{x}\left(v_{y}\right)\left|\psi_{c\mathbf{k}}\right\rangle }{E_{c\mathbf{k}}-E_{v\mathbf{k}}},
\end{equation}
are related to the energy difference between the conduction and valence
bands, $E_{c\mathbf{k}}-E_{v\mathbf{k}}$, where $v$ and $c$ refer
to the valence and conduction bands, respectively.

\begin{figure}
\begin{centering}
\includegraphics[scale=0.5]{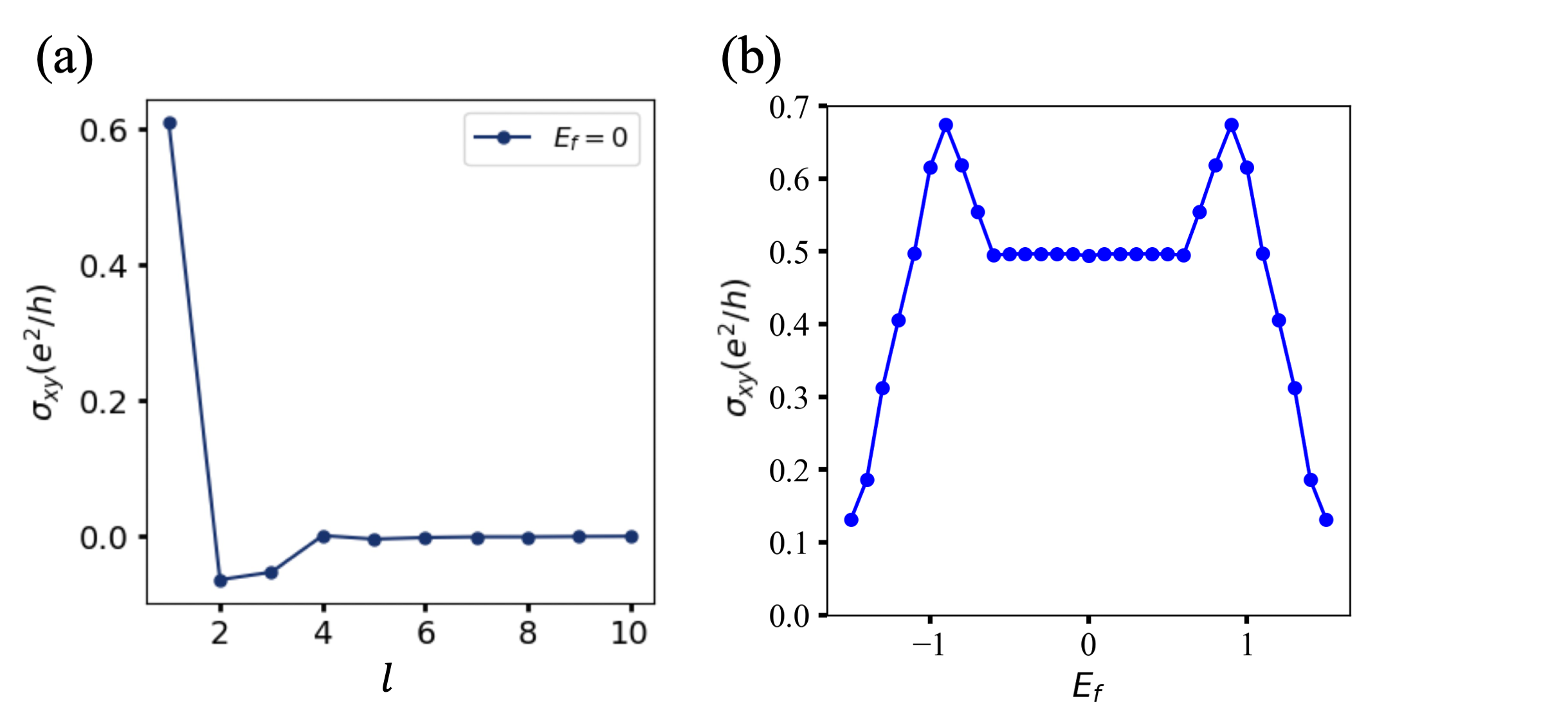}
\par\end{centering}
\caption{\label{fig:6}(a) Layer-resolved Hall conductance  $\sigma_{xy}(l)$
as a function of
the layer index $l$. (b) Surface Hall conductance $\sigma_{xy}^{\text{surf}}$ 
as a function of the Fermi energy $E_f$.}
\end{figure}

Figure \ref{fig:6}(a) illustrates the layer-resolved Hall conductance distribution at \(E_{f} = 0\). Due to the introduction of altermagnetism at \(l = 1\), which breaks time-reversal symmetry, the Hall conductance becomes non-zero in the first three layers, while the local anomalous Hall conductance (AHC) becomes nearly zero for \(l > 3\). It is important to note that the layer-resolved Hall conductance distribution is largely independent of the Fermi level, which is why we focus on the case of \(E_{f} = 0\) for clarity. The surface Chern number is determined by summing the local Chern markers from the first few layers \citep{varnava2018surfaces,gu2021spectral}. Therefore, we define the surface AHC as \(\sigma_{xy}^{\text{surf}} = \sum_{l=1}^{3} \sigma_{xy}(l)\). Under this definition, Fig. \ref{fig:6}(a) shows that the surface Hall conductance at \(E_{f} = 0\) approaches \(e^{2}/2h\).

Figure \ref{fig:6}(b) presents the variation of the surface AHC \(\sigma_{xy}^{\text{surf}}\) with the Fermi level, revealing a plateau at \(e^{2}/2h\) in the central energy region. This plateau occurs where the Fermi level intersects only a single Dirac cone, signifying the persistence of the half-quantized Hall conductance across a range of energies. The origin of this Fermi level-independent half-quantization lies in the parity invariance of the Fermi surface, ensuring the robustness of the topological behavior induced by altermagnetism, even as the Fermi level varies\citep{zou2023halfquantized}. 

\section{Discussion and conclusion}

Before concluding, we will discuss the differences between realizing
the parity anomaly in weak 3D TIs and in strong 3D TIs. 
Although lattice regularization is absent in strong 3D TIs, 
field theories with anomalies can describe the boundary
states of topological phases in a higher spatial dimension. 
This explains
why 2+1D massless Dirac fermions with parity anomaly can exist on
the surface of a strong 3D TI\citep{fu2007topological}. 
The topological surface 
states avoid violating the fermion doubling theorem because their
ultraviolet completion resides in the bulk. This manifests in the
behavior of surface states on the top and bottom surfaces of a strong 3D
TI, where at higher momentum, the surface states gradually evolve
into bulk states. This evolution enables coupling between the top
and bottom surface states at high momentum\citep{zou2023halfquantized,wang2024signature}.
Consequently, in the case of a semi-magnetic TI, when time-reversal
symmetry is broken on one surface, the other surface can still support
a massless linear dispersion at low energies, while incorporating
a term that breaks time-reversal symmetry at higher energy, leading
to the emergence of half quantized Hall conductance\citep{zou2023halfquantized}.

For weak 3D TIs, the strong \(\mathbb{Z}_2\) invariant is equal to 0, leading to a fermion doubling problem on the surface. Due to lattice regularization, a
gapless Dirac cone with chirality opposite to that at the $\Gamma$
point emerges at the $Y$ point. This makes it impossible to realize a parity anomaly
through conventional mechanisms used in 3D TIs, such as introducing
ferromagnetism on one surface.  

In this work, we demonstrate that the introduction of altermagnetism allows weak topological insulators to achieve the realization of the parity anomaly. It preserves time-reversal symmetry at low
energies near the $\Gamma$ point while breaking it at the $Y$ point,
where the gapless Dirac cone present can be understood as a result of lattice regularization.
Here, altermagnetism replaces the role of lattice regularization,
thereby avoiding the fermion doubling problem. This enables the surface
to support a single gapless Dirac cone, resulting in a half-quantized
Hall conductance due to the parity anomaly.

Our model can be regarded as an experimentally feasible adaptation
of the Haldane model. Like the Haldane model, our approach does not
require a net magnetic field to realize the parity anomaly. However,
unlike the Haldane model, which demands fine-tuning and relies on
staggered magnetic fluxes that are challenging to achieve experimentally,
our approach involves introducing altermagnetism on the surface of
a weak TI, which can be achieved experimentally through proximity
effects and is significantly easier to implement.

The combination of altermagnetism and weak TIs is particularly promising
because both types of materials have already been explored experimentally.
Altermagnetism has been observed in materials such as MnTe\citep{krempasky2024altermagnetic,osumi2024observation}, CrSb\citep{reimers2024directobservation}, $\textrm{Ru\ensuremath{O_{2}}}$\citep{fedchenko2024observation},
while several materials, including $\textrm{B\ensuremath{i_{4}I_{4}}}$\citep{noguchi2019aweak}, $\textrm{ZrT\ensuremath{e_{5}}}$\citep{zhang2021observation}, $\textrm{B\ensuremath{i_{4}}B\ensuremath{r_{2}I_{2}}}$\citep{zhong2023towards},
have been experimentally confirmed as weak TIs. Notably, a recent experiment has demonstrated the stability of surface states in weak topological insulators\citep{PhysRevLett.133.086602}. Given the experimental
verification of these key ingredients, we believe that the approach
presented in this paper is experimentally feasible.

\section*{acknowledgements}

Y.-H. W. is grateful to  L.-D. Z. for fruitful discussions.
This work was financially supported by NSF-China (Grants No. 11921005 and No. 12374034),
the Innovation Program for Quantum Science and
Technology (No. 2021ZD0302403), and the Strategic
priority Research Program of the Chinese Academy of
Sciences (Grant No. XDB28000000). 
We also acknowledge the High-performance Computing Platform of Peking University for providing computational
resources.

\bibliography{main}

\end{document}